# DISCUSSION OF: STATISTICAL ANALYSIS OF AN ARCHAEOLOGICAL FIND

By Donald L. Bentley

*Pomona College*

I begin this discussion by quoting a Mosaic law. This is not one that can be found in the Torah, but I know it to be authentic because I heard it from the mouth of Moses himself. The law is, "Statistics is the umpire of the sciences." This law was told me by Lincoln Moses, one of the top applied statisticians of the twentieth century, a real craftsman with data and a master of the application of statistics and statistical reasoning.

To appreciate this law, one needs to make the distinction between theoretical and applied statistics which I would like to illustrate with an example. A number of years ago, Joe Gani gave a talk to a group of statisticians. During the talk, which included a discussion of Fisher's work on predicting the number of species of butterflies in Malaysia, he made an aside remark that perhaps the same method could be used to determine the number of words in a person's vocabulary. The problem, posed in a general mathematical context, resulted in the well-known paper by Efron and Thisted (1976) titled "Estimating the number of unknown species: How many words did Shakespeare know?" And while the authors used the vocabulary framework for the structure of their research, there was no specific application in mind. The intended audience was the statistics community, not a group of Shakespearean scholars. I consider this an example of theoretical statistics.

Later, Thisted happened upon an article reporting that a newly discovered poem could well have been written by Shakespeare. Thisted and Efron (1987) set about modifying their previous results so the assumptions and methodology met the requirements necessary for applying them to the question of whether there was reason to believe that this poem had not been written by Shakespeare. The statistics now served as a tool for the primary purpose, not an end in itself. This is applied statistics. I believe the Mosaic Law, "Statistics is the umpire of the sciences," is directed toward statisticians working on applied problems. Because the problem which motivated the paper under discussion is an application, it seems appropriate to consider how well the statistics served as umpire in the research.[1]

I believe there are five phases of a research project in which an applied statistician should be intimately involved, and the properly prepared statistician will feel comfortable participating in each of them [Bentley (1996)]. They are: (1) determining the question, (2) designing the experiment, (3) gathering and validating the data, (4) analyzing the data, and (5) communicating the results. There is a need for statistical reasoning to be applied at each phase of the project, not just during the analysis of the data. The statistician should be involved from the very beginning, which is the forming of the question the research team will attempt to answer. It is the statistician's responsibility to make sure everyone on the team understands the question and is in agreement that it is the appropriate question for the project. Moreover, equally as important is that the statistician makes sure everyone is in agreement with the assumptions behind the question. This includes making sure these assumptions form a consistent set. There should be ample evidence that they can be reasonably accepted by the community of scholars in the field of the application. The statistician, though not an expert in the substantive field, should still feel comfortable with the arguments being used by the research team members to justify the assumptions. And the assumptions should not preclude being able to answer the agreed-upon question; they must be consistent with the question. Keep in mind that a good umpire is not required to be highly skilled in playing the game, but must have a good knowledge of the rules by which the game is to be played and have had enough exposure to the game to be able to detect when the rules are being violated.

The abstract of the paper under discussion states, "An approach is proposed for measuring the 'surprisingness' of the observed outcome relative to a 'hypothesis' that the tombsite belonged to the NT family." This is followed in Section 1, the Introduction and Summary, with the statement, "Since names such as Yehosef, Marya, Yeshua, etc., were not uncommon during the era in which such burials took place, the task of assessing whether or not these ossuaries might be those of the New Testament (NT) family is not straight forward. [...] One purpose of this article is to contribute toward such efforts by developing statistical methods for assessing evidence for and against a 'hypothesis' that this tomb belonged to the family of the historical Jesus."

---

[1] The results of the research were first released by Discovery Channel at a press conference held at the New York City library on February 26th and then to the general public on March 4th, 2007 in the broadcast of their documentary "The Lost Tomb of Jesus," produced by James Cameron and directed by Simcha Jacobovici.

The analysis addresses this question under a given set of assumptions. Therefore we can only accept the conclusion if we are willing to accept the assumptions. As umpires, we need to know exactly what these assumptions are and the consequences these assumptions impose upon any inferences that are drawn from the analysis. Several times in the paper, the author refers to a "historical viewpoint" and "historical assumptions." I was unable to find a definition of "historical," but based on the assumptions and arguments presented in the text, I have been led to believe that by historical viewpoint is meant a strictly literal translation of the Gospels with one exception. The final paragraph of Section 1 states, "We remark that in assessing the evidence in any way, it is essential to adopt a strictly *historical* viewpoint, and thus to set aside considerations that a NT tombsite cannot exist. In fact, Jewish ritual observances prevalent at the time are entirely consistent with the possible existence of such a tomb." This wording is equivalent to assuming a New Testament family tomb might exist, that it is possible there could be a family tomb which might even contain the remains of Jesus of Nazareth.

However, this is not the assumption that was used in the analysis. Instead, the analysis is conditioned on the assumption that, with probability one, there did exist such a tomb. And even beyond that is the assumption that this Jesus family tomb was in the vicinity of Jerusalem with probability one. In Section 14, the author states, "We are in fact now in a position to carry out a particular *hypothesis test*: Here $H_0$ is the hypothesis that *all* 1,100 tombs in the vicinity of Jerusalem arose under random assignments of names, and $H_1$ is the hypothesis that *one unspecified one among these 1,100 tombs is that of the NT family.*" However, the calculations do not seem to allow for an *a priori* probability greater than zero that there does not exist a family tomb in Jerusalem.

This raises a question. Even if there were a tomb for the family of Jesus of Nazareth, why would it be in Jerusalem with probability one? Why would it not be in the Galilee around Nazareth or Capernaum which was the base of Jesus' ministry? James Tabor is acknowledged in the paper as being a New Testament expert and in particular for his involvement in formulating the "A Priori Hypotheses," the assumptions that were used in the analysis. In checking Tabor's new book, *The Jesus Dynasty*, for an answer to this question one finds a picture on page 239 of Tabor kneeling on the Tsaft grave which is located in the Galilee, near Capernaum [Tabor (2006)]. In the associated text, Tabor describes a rabbinic tradition which identifies the grave as the tomb of Jesus of Nazareth, and Tabor presents an argument, based on the Gospels, as to why Jesus' body might have been returned to that area for burial. From this discussion, it appears that in 2006 Tabor did not believe the assumption that Jesus was buried in a family tomb in Jerusalem with probability one. However, he was willing to accept that assumption in establishing the "A Priori Hypotheses" for this analysis.

Another assumption which is not stated explicitly in the paper is incorporated into the estimate of the numbers of ossuaries and tomb sites in the Jerusalem area. Section 8 begins, "We require estimates of the size of the relevant population of Jerusalem and of the number of ossuary burials that took place overall." Based upon an estimate of the number of *residents* in Jerusalem during the period of ossuary burials, the researchers came up with an estimate of about 6,600 inscribed ossuaries which, when assuming a configuration of four males and two females per tombsite as in the Talpiyot site, led to an estimate of 1,100 tombsites in the Jerusalem area which is the figure used in the analysis. But this estimate excludes those persons who were not residents of Jerusalem yet might have been buried there in family tombs. Jesus of Nazareth, for example, was not included, nor was Mary of Magdala, nor was Joseph from Arimathaea whom the author argues arranged for the tomb in which Jesus was first buried.

There are two important factors the author ignored in estimating the number of ossuary burials that took place in the Jerusalem vicinity. First, it was sacred tradition that Jews make a pilgrimage to the Temple in Jerusalem three times a year, at the three important Festivals. One of these was the Passover when Jesus was crucified. Based on extra-biblical sources, scholars estimate that the number of people in Jerusalem increased by between fourfold and tenfold during these Festival periods. In other words, the number of people in Jerusalem at the time of Jesus' death would have been between 120,000 and 300,000, not 30,000. In fact, E. P. Sanders (1992) estimates the number to have been between 300,000 and 500,000. Further, many Jews then, even as today, desired to be buried in Jerusalem. There would have been nothing to prevent a family from bringing the bones of a family member from afar to place in an ossuary in a family tomb in Jerusalem, just as the bones of Joseph were carried out of Egypt. These omissions cause the assumptions used in calculating the number of tombsites in Jerusalem to be inconsistent with other assumptions used in the statistical analysis. In particular, assumptions A.5 and A.7 of the paper are inconsistent.

Among the greatest sources of controversy arising from *The Lost Tomb of Jesus* documentary is the Mariamene inscription, and the ossuary's "relevance" to Mary of Magdala.[2] At issue is whether the inscription refers to one or two persons. Assumption A.7 of the paper states, "We assume that the full inscription Mariamenou [η] Mara refers to a single individual and represents the most appropriate specific appellation for Mary Magdalene amongst those known...." The impact this assumption has on the results of the analysis is acknowledged by the author in the paper's final paragraph: "Among the various assumptions made, perhaps the one that most "drives" our analysis in the direction of 'significance' is the extraordinary inscription Mariamenou [$\eta$] Mara." In describing Ossuary #1, the author states, "Rahmani (1994), pages 14, 222, reads the inscription as follows: "The stroke between the $\upsilon$ of the first and the $\mu$ of the second name probably represents an $\eta$, standing here for the usual $\eta$ $\kappa\alpha\iota$... used in the case of double names..." and he posits that the second name is a contracted form [not a contraction] of 'Martha' leading to the reading 'Mariamene [diminutive] who is also called Mara'." The author then adds that it is Rahmani's reading which was adopted in determining the relevance of the ossuary for the analysis with the justification that it "was accepted by Kloner (1996) and has been corroborated by others [without reference] in the field." It should be noted that Kloner is not an epigrapher but rather one of the first archaeologists into the tomb in 1980.

---

[2]See Bovon (2007) concerning his opinion of the inappropriateness of the appellation of Mariamenou for Mary Magdalene in the first century CE, as compared to the impression given of his expert opinion in the documentary. Also more generally, Shanks (2007).

In determining how to evaluate the inscriptions of the Yoseh and Marya ossuaries the author rejected Rahmani's interpretation. Near the end of Section 2 the author states, "Rahmani surmised that the similarities between ossuaries #5 [Yoseh] and #6 [Marya] and their inscriptions, both coming from the same tomb, may indicate that Yoseh and Marya were the parents of Yeshua and the grandparents of Yehuda." In the attached footnote 5 the author adds, "If this interpretation is correct, the tombsite cannot be that of the NT family. However Rahmani does not follow up with any explanation for the messy nature of the inscription of Ossuary #4." Rahmani's suggested interpretation is treated by the author in assumption A.4 which states, "We assume that the ossuary inscribed 'Yehuda son of Yeshua' can be explained and may be discarded in our analysis." The umpire in me has to ask a question. Why should we be willing to accept with probability one Rahmani's interpretation of the "Mariamenou" inscription which supports the desired conclusion of the researchers although Rahmani only claims it is "*probable*," yet be willing to reject with probability one an interpretation which Rahmani states "*may be indicated*" when the interpretation would invalidate their theory? It should be pointed out that these decisions were made a posteriori by those who formulated the eight "A Priori Hypotheses" and the nine "Assumptions," after they had seen the data.

The other side in this debate claims that the Mariamenou inscription should be interpreted to refer to two women buried at different times, one with the name Mariame and the other the name Mara. It should be noted that it was quite common to have multiple burials in a single ossuary. Stephen Pfann (2007) provides very convincing evidence for this position. He notes that the "mark" between the first and second names, which the author accepted as representing $\eta$ $\kappa\alpha\iota$, is in fact just a scratch made with a different tool than the tools used for the first and second names, and in fact the two names were carved with different tools. Further, Pfann suggests the reader compare the handwriting between the letters which form Mariame and the letters which form Mara. The first letter in each group is a capital $mu(M)$, the second an $alpha(\alpha)$, and the third a $rho(\rho)$. He points out that the first and second letters of the first name are each printed with two strokes and the third, the *rho*, is one continuous stroke. On the other hand, the first two letters of the second name are each made with a single continuous stroke, while the *rho* is formed with two clearly distinct strokes. I am not an expert in ancient Greek writing, but as an umpire I would need some convincing evidence before I would be willing to accept the assumption that the first and last words in the inscription were done by the same person and at the same time. And if they were not done at the same time, most likely the Mara inscription would be the name of a second person buried in the ossuary at a later date.

Jurgen Zangenberg (2007), a biblical archaeologist at Leiden University, addressed the use of the assumption that the Mara on the Mariamene inscription should cause it to be read as Mariamenou the Master with a rhetorical question, asking why one should assume that "Mara" here must be an honorary title unless one wishes to "prove" that which is already assumed to be known. What he is saying is that it is possible to prove anything that you assume to be true. It is the job of the statistician, the umpire, to make sure the assumptions do not provide for this type of circular logic.

The examples I have given deal only with the first three of the five points where an applied statistician should have contributed to this research project: the point of determining the question the research was to answer and then making sure the assumptions used in the analysis were valid and also were consistent with the question, designing the experiment to gather appropriate data, and then collecting and validating the data. The concerns I raised above about the assumptions are just a few from many that I had as I read through the paper. The author, in numerous places throughout the paper, makes comments such as the following which appears in Section 1, the Introduction and Summary. "Our computations were carried out under a specific set of assumptions...." And in the first paragraph of Section 13, A Statistical Analysis, we find the statement, "The assumptions A.1–A.9 under which we carried out our analysis are by no means universally agreed upon. Furthermore, the failure of any one of them can be expected to impact significantly upon the results of the analysis." There is a footnote to this statement which warrants particular attention. The footnote states, "These assumptions were proposed by S. Jacobovici, except for A.6 & A.9 which are due to the author."

Simcha Jacobovici is coauthor of the book, *The Jesus Family Tomb*[3] [Jacobovici and Pellegrino (2007)], as well as executive director of the Discovery Channel's documentary, *The Lost Tomb of Jesus.* At the press conference held on February 26, 2007 at the New York public library, Jane Ruth, the president and general manager of the Discovery Channel, referred to the subject of the documentary as, "what might be one of the most important archaeological finds in human history." After introductory remarks by Ms. Ruth and the documentary's producer, James Cameron, the podium was turned over to Jacobovici to provide the facts about the find. Following his presentation and prior to allowing questions from the media that were directed to the panel of experts, Jacobovici made the following statement. "Before I turn it over to the experts, because I have to say again, I'm not going to say I'm not an expert. I've seen a lot of internet buzz on this. I am an expert. My expertise is investigative journalism. I'm not an archaeologist. I'm not a DNA expert. I'm not a statistician. I'm a filmmaker and a journalist."

During the questioning by the media, the statistician who is the author of the paper under discussion had the opportunity to speak. He began, "The obvious needs to be stated; that I'm not a biblical scholar, not a historical scholar. I'm just a numbers guy.... As a statistician, I do the calculations based on assumptions given to me *by the subject matter experts, in this case the historical biblical scholars.*"

The above quotes from the press conference raise two concerns. First, the responsibility of the statistician as an umpire of the science cannot be fulfilled by just accepting a given set of assumptions. They must be checked to make sure they meet certain standards as identified at the beginning of this discussion. But beyond that, the author did not even satisfy his own standards of using assumptions provided by "subject matter experts." By Jacobovici's own admission, his areas of expertise are filmmaking and journalism, and not the substantive fields of the complex history and archaeology of first century Judaism. Yet, as stated in both footnote 32 and the acknowledgments of the paper, it was Jacobovici who provided the assumptions which served as the basis for the statistical analysis.

---

[3]It should be noted that on page 114 of *The Jesus Family Tomb* there is a footnote that reads, "As of this writing, Feuerverger's paper has been submitted to a leading American statistical journal and is being peer-reviewed." It must be emphasized that this peer review is being performed by statisticians who are not experts in archaeology, biblical history, or epigraphy. This review is therefore restricted to the statistical reasoning involved in the research, and should not be used as a reference for assuming the validity of the assumptions which served as the bases for the statistical analysis. The review of archaeological and biblical issues must be performed by the appropriate subject matter experts.

Unfortunately, the lack of acceptance by the archaeological community, not to mention biblical scholars, of many of the assumptions used in this analysis is being recognized as a problem attributable to the field of statistics. As an applied statistician who is attempting to introduce more statistical reasoning into biblical archaeology, perhaps the most distressing comments I have read concerning the *Jesus Family Tomb* project were made by Sandra Scham (2007) in an article titled "The 'Jesus Tomb' on TV" which appeared in *Archaeology*, a journal she edits which is a publication of the highly regarded Archaeological Institute of America. Noting the reception this project has received among archaeologists and biblical scholars, she wrote, "At one time archaeologists loved statistics, happily performing complex regression and cluster analyses on our data and spitting out conclusions from our computers that, likely, proved the conjectures we had begun with. In the last two decades, however, we have begun to question these facile validations of our common sense. The problem is with the data. The methods may be perfectly suited to a world in which a representative sample, normal distribution, or even an idea of what the population in question might be, is possible. Archaeological evidence is precisely the opposite. We do not, in point of fact, know any of these things. In the words of one former statistically enthralled antiquarian, 'Even when the odds were good, we knew the goods were odd.'"

In my role as an umpire of "The Statistical Analysis of an Archaeological Find," I find myself in agreement with Ms. Scham.

Department of Mathematics
Pomona College
Claremont, California 91711
USA
E-mail: dbentley@pomona.edu